\documentstyle[prd,aps,eqsecnum,amsfonts]{revtex}
\begin{document}
\draft


\title{Quantum scalar field in $D$-dimensional static black hole
space-times.}
\author{Daniele Binosi\footnote{email: binosi@alpha.science.unitn.it}}
\address{Dipartimento di Fisica, Universit\`a di Trento, Italia}
\author{Sergio Zerbini \footnote{email: zerbini@science.unitn.it}}
\address{Dipartimento di Fisica, Universit\`a di Trento \\ 
and Istituto Nazionale di Fisica Nucleare,\\
Gruppo Collegato di Trento, Italia}
\date{October 30, 1998}
\maketitle

\begin{abstract}

An Euclidean approach for investigating quantum aspects of a scalar field
living on a class of $D$-dimensional static black hole space-times,
including the extremal ones, is reviewed.
The method makes use of a near horizon approximation of the metric
and~\mbox{$\zeta$-function} formalism for
evaluating  the partition function and the 
expectation value of the field fluctuations~$\langle\phi^2(x)\rangle$.
After a review of the non-extreme black hole case, the extreme one is
considered in some details. In this case, there is no conical singularity,
but the finite imaginary time compactification introduces a cusp
singularity. It is found that the~\mbox{$\zeta$-function} 
regularized partition function can be defined, and the quantum fluctuations 
are finite on the horizon,
as soon as the cusp singularity is absent, and the corresponding
temperature is $T=0$.

\end{abstract}


\section{Introduction.}

The issue of determining the equilibrium (Unruh-Hawking) temperature of a
black hole, is important. In fact, one can extract 
thermodynamical informations from its knowledge: for example the
Bekenstein-Hawking entropy (\emph{i.e.} the tree-level contribution to the
entropy) can be defined as the response of the free energy of the black
hole
to the change of this equilibrium temperature. Furthermore, it defines 
the admissible temperatures of thermal states of free scalar fields in a
static and globally hyperbolic space-time region with horizons.

As is well known, there exist several methods for evaluating the possible
equilibrium temperature of a stationary black hole. Whithin the simplest
of these methods, 
one  has to make a Wick rotation of the time coordinate (passing in this
way to the Euclidean time~$\tau=it$), and eliminate all the metric
(conical) singularities connected to the horizon by an opportune 
choice of the time periodicity~$\beta_M$ \cite{hawking1}. 
Then, one has to impose the KMS
condition for thermal states~\cite{kubo,haag}, \emph{i.e.} to impose the
periodicity condition on the imaginary time dependence of the thermal
Wightman functions, and interprete the common period~$\beta_T$ as the
inverse
of the temperature~$T$ of the state.\\
Although this procedure determines the correct Unruh-Hawking temperature
in
the case of a non-extreme black hole, it does not apply to the extreme
case
(for example to the case of an extreme Reissner-Nordstr\"om black hole),
since one is unable to determine the time periodicity of the
manifold~$\beta_M$.

Later, a more sophisticated Lorentzian  method was introduced 
in~\cite{haag1} and
successively developed in~\cite{kaywald} and~\cite{hessling,moretti0}.
Without entering in the  details of this approach, we only recall that
the method is connected to the well known Hadamard expansion of the
two-point
Green functions in a curved background and in the limit of coincidence of
the
arguments. Basically in~\cite{haag1} was proved that assuming fairly
standard
axioms of quantum (quasi-free) field theory (such as local definitness and
local stability in a stationary space-time region), then, when the
distance
between the two arguments vanishes, the thermal Wightman functions in the
interior of this region will transform into non-thermal and massless
Wightman
functions in Minkowski space-time.
This point coincidence behavior of the Wightman functions, must hold for
any
phisically sensible state (thermal or not), and, in the case that the
space-time region one is dealing with is just a part of the whole manifold
separeted by event horizons, it must hold on the horizons. This constraint
actually selects the correct temperatures~$T=\beta_T^{-1}=\beta_M^{-1}$,
in
the case of Rindler and Schwarzschild space-times.
Then these results. have been generalized in~\cite{kaywald}
to a large class of space-times admitting an appropiate reflection
isometry.

Both Haag's method~\cite{haag1} and Kay-Wald approach~\cite{kaywald},
which, as they stand, work only for space-times with an intersection
between past and future horizons, were extended
in~\cite{hessling,moretti0}
for working in more physical situations than the ethernal black holes one,
including the extreme case.\\
However, all of these ``Lorentzian'' methods involve a certain  amount of 
calculations (for example the procedure developed in~\cite{moretti0},
requires
the evaluation of all the possible geodesics which start from the
horizon);
for this fact, even if it is not difficult to foresee a
possible generalization to, say, $D$-dimensional extreme black holes,
the concrete computation does not appear an easy task.\\
In this paper, making use of Euclidean methods, we would like to obtain
some
informations about the equilibrium temperature of a class of static,
$D$-dimensional black holes, evaluating quantities such as the field
fluctuations and the one-loop partition function. The inverse of the
temperature is formally introduced as the period of the compactified
imaginary
time~$0\leq\tau\leq\beta$.

In order to deal with explicit calculations, we will make use of a near
horizon approximation of the metric. This approximation may also be
justified
observing that only near the horizon interesting physical effects are
supposed to be relevant.

First, the case of non-extreme black holes is reconsidered. Here, as is
well
known, a conical singularity is present. We will show that its presence
leads
to divergences of the field fluctuations on the horizon. However in this
case
it is known that as soon as the smoothness of the manifold is required,
the
Hawking temperature, as well as the absence of the divergences, is
recovered.

The analisysis then extended to the extreme black holes case. In this
case,
no conical singularity is present, but the compactification of the
imaginary
time and the related periodic identification, induces an isometry
containing
parabolic elements (translation in $\tau$), so that a cusp singularity
appears. Its presence leads to the following features:
\begin{enumerate}
\item The field fluctuation have divergences on the
horizon;
\item The global~\mbox{$\zeta$-function}, besides the
horizon divergences,
does not exist, and requires a further regularization.
\end{enumerate}
These undesired features disappear as soon as the imaginary time period is
taken to be~$\infty$, namely the associated temperature is to be~$T=0$, 
in agreement with the four-dimensional results obtained in 
\cite{anderson,moretti0}.

The main objection to the approach proposed here could be the use of a
near
horizon approximation of the metric. However, we stress that in
\cite{haag1} only the limit form of the metric near the horizon was use in
order to obtain the Unruh-Hawking temperature; moreover also the results
in \cite{anderson,moretti0}  were derived in a near horizon approximation 
contest.\\

The paper is organized as follows.
In Sec.~II we review the evaluation of the field fluctuations within 
the~\mbox{$\zeta$-function} regularization procedure, while in Sec.~III we
will derive a
near horizon approximation of the generic line element describing
a non-extreme and an extreme black hole. Then in Sec.~IV and~V we will
discuss in detail these two cases, taking advantage of the approximation
done. The paper ends with some concluding remarks in Sec.~VI.

\section{Evaluating the field fluctuations with
the~\mbox{$\zeta$-function}
procedure.}\label{secfieldfluctuations}

As  anticipated in the Introduction, we need to evaluate the
expectation
value of the field fluctuations~$\langle\phi^2(x)\rangle$, in the
framework of
the~\mbox{$\zeta$-function} regularization procedure.
Within this approach (see~\cite{morettiultimo} for an exhaustive
discussion)
one has~\cite{cognola0,iellici0}
\begin{eqnarray}
\langle\phi^2(x)\rangle & = & -\frac2{\sqrt{g(x)}}
\frac{\delta S_{\mathrm{eff}}}{\delta J(x)}_{\vert J(x)\equiv 0} \nonumber
\\
& = & \frac1{\sqrt{g(x)}}\frac d{ds}\left[\frac{\delta\zeta(s\vert
A\ell^2)}
{\delta J(x)}\right]_{\vert s=J(x)=0},
\end{eqnarray}
where~$J(x)$ is a classical source, and~$\ell$ is the usual arbitrary
parameter (with the dimension of mass$^{-1}$) necessary from dimensional
considerations.\\
By a direct calculation, it follows that  the $\zeta$-regularized 
field fluctuations turns out to be
\begin{equation}
\langle\phi^2(x)\rangle_{\mathrm{ren}}=\ell^2\frac d{ds}\left[s
\zeta(s+1;x\vert A\ell^2)\right]_{\vert s=0},
\label{firstevenfieldfluctuations}
\end{equation}
with the~\mbox{$\zeta$-function} evaluated when the source~$J(x)$
vanishes.\\
By making use of the the Laurent expansion of~$\zeta(s+1;x\vert A\ell^2)$,
and extracting from the~\mbox{$\zeta$-function} the~$\ell^2$ dependence,
we can rewrite~(\ref{firstevenfieldfluctuations}) as
\begin{equation}
\langle\phi^2(x)\rangle_{\mathrm{ren}}=\lim_{s\to0}\left[
\zeta(s+1;x\vert A)-\frac1s{\mathrm{Res}}\,
\zeta(s+1;x\vert A)-{\mathrm{Res}}\,\zeta(s+1;x\vert A)\ln\ell^2\right].
\label{evenfieldfluctuations}
\end{equation}
Notice that when the manifold is smooth, the meromorphic 
structure of the~\mbox{$\zeta$-function} is known (Seeley's Theorem). In
particular 
for a differential elliptic operator of the second order (Laplacian) one
has 
\begin{equation}
\Gamma(z)\zeta(z;\vec x\vert L_N)=\sum_{r=0}^{\infty}\frac{A_r(\vec
x\vert L_N)}{z+r-\frac N2}+\textrm{analytic part},
\qquad A_r(\vec x\vert L_N)=\frac{a_r(\vec x\vert L_N)}{(4\pi)^{\frac
N2}}.
\label{transversezf1}
\end{equation}
The spectral coefficients~$a_r(\vec x\vert L_N)$ are computable functions
known as the Seeley-de Witt coefficients.\\
As a consequence, if the dimension of the smooth manifold is odd, 
the~\mbox{$\zeta$-function} is regular at~$z=1$~($s=0$) and the dependence
on the scale
parameter~$\ell$ disappears and one gets
\begin{equation}
\langle\phi^2(x)\rangle_{\mathrm{ren}}=
\zeta(1;x\vert A).
\label{evenfieldfluctuations1}
\end{equation}
On the other hand, if the dimension is even,
there is a simple pole at~$z=1$, and the~$\ell$ ambiguity will be present. 

\section{Near Horizon Approximation of the Metric.}

The metric for a general static spherically symmetric $D$-dimensional
space-time, analytically continued into the Euclidean space, reads
\begin{equation}
ds^2=f(r)d\tau^2+\frac1{h(r)}dr^2+r^2d\Sigma_N^2, \qquad
x=(\tau,r,\vec{x}),
\end{equation}
where~$\tau=it$ is the Euclidean time,~$f$ and~$h$ are arbitrary functions
of~$r$ (which are constant in the ~$r\to\infty$ limit, if the space-time
has
to be asimptotically flat), and~$d\Sigma_N^2$ represent the line element
of a
smooth N-dimensional (transverse) manifold without boundary
($\vec x$ are the transverse coordinates).

For this metric representing a black hole, one demands the presence,
at~$r=r_+$, of a zero in both~$f$ and~$h$, so that, according to the 
nature of this zero, one finds the following two interesting cases.

\subsection{Non-extremal case.} 

In the case of a non-extremal black hole, one has a simple zero
at~$r=r_+$,
so that the functions~$f$ and~$h$ can be expanded as \cite{beke96}
\begin{equation}
f(r)\simeq f'(r_+)(r-r_+), \qquad h(r)\simeq h'(r_+)(r-r_+).
\end{equation}
Thus, after changing to the coordinates~$(\rho,\theta,\vec{x})$ by means
of
\begin{equation}
\rho^2=\frac4{h'(r_+)}(r-r_+), \qquad
\theta=\frac12\sqrt{f'(r_+)h'(r_+)}\tau,
\label{simplezerocoords}
\end{equation}
the geometry near the event horizon is described by the approximated line
element
\begin{equation}
ds^2\simeq d\rho^2+\rho^2d\theta^2+r^2_+d\Sigma^2_N.
\label{po}
\end{equation}

We may generalize the argument to black hole solutions in semiclassical 
gravity. In this case, near the horizon, one has 
\begin{equation}
f(r)\simeq C_f(r-r_+)^{c_1}, \qquad h(r)\simeq C_h(r-r_+)^{c_2}.
\end{equation}
with the constants $C_f >0$, $C_h >0$, $c_1 >0$, $0 <c_2 <2$. 
Here $c_1$ may be less than one, and the first 
derivative may not exist at the horizon. However, if $c_1=2-c_2$, it is 
easy to show that by means of the following coordinates transformation
\begin{equation}
(r-r_+)^{c_1/2}=\frac{c_1}2\sqrt{C_h} \rho, \qquad 
\theta=\frac{c_1}{2} \sqrt{C_hC_f} \tau.
\end{equation}
the line black hole element reduces again to the line element (\ref{po}). 
For example, the previous case  corresponds to $c_1=c_2=1$, and  very 
recently, in \cite{paul} the case $c_1=1/2$ and
$c_2=3/2$  have been considered; in any case notice that  
since $c_2 < 2 $, the proper radial distance to the horizon is finite.

Now, finite temperature effects are assumed to arise when the Euclidean
time
$\tau$ (correspondingly $\theta$) is compactified requiring 
$0\leq\tau\leq\beta$ 
($0\leq\theta\leq\gamma$), with $\beta$ the inverse of the temperature.
So,
for arbitrary $\beta$ ($\gamma$), the manifold ${\mathcal{M}}^D$ shows, 
near the
horizon, the topology of ${\mathcal{C}}_\gamma\times\Sigma^N$,
${\mathcal{C}}_\gamma$ being the simple two-dimensional flat cone with 
deficit angle $2\pi-\gamma$.

In such a space-time, one usually  determines the temperature of the
black hole, by requiring the absence of the conical singularity
\cite{hawking1}: the manifold, in fact, is not smooth, showing a conical
singularity at $\rho=0$ unless $\gamma=2\pi$. In this way the temperature
is
found to be
\begin{equation}
T=\frac{\sqrt{f'(r_+)h'(r_+)}}{4\pi},\qquad 
T=\frac{c_1}{4\pi}\sqrt{C_hC_f},
\end{equation}
respectively, which are the  Unruh-Hawking temperatures of the black
holes. 

We will  show that
$\gamma=2\pi$ is the only possible requirements for having a well-behaved
$\langle\phi^2(x)\rangle$ on the horizon.
Now, finite temperature effects are assumed to arise when the Euclidean
time~$\tau$ (correspondingly~$\theta$) is compactified
requiring~$0\leq\tau\leq\beta$ 
($0\leq\theta\leq\gamma$), with~$\beta$ the inverse of the temperature.
So,
for arbitrary~$\beta$ ($\gamma$), the manifold~${\mathcal{M}}^D$ shows, 
near the horizon, the topology
of~${\mathcal{C}}_\gamma\times\Sigma^N$,~${\mathcal{C}}_\gamma$ being
the simple two-dimensional flat cone with deficit angle~$2\pi-\gamma$.

\subsection{Exteme case.}

In the case of an extreme black hole, one has a double zero at~$r=r_+$, so
that the behavior of~$f$ and~$h$ near the horizon, is \cite{beke96}
\begin{equation}
f(r)\simeq\frac12f''(r_+)(r-r_+)^2, \qquad h(r)
\simeq\frac12h''(r_+)(r-r_+)^2.
\end{equation}
Thus, if we define the new coordinates~$(\rho,\theta,\vec{x})$ by means of
\begin{equation}
\rho=\sqrt{\frac2{f''(r_+)}}(r-r_+)^{-1}, \qquad
\theta=\sqrt{\frac{h''(r_+)}2}\tau=\frac{\tau}b,
\label{doublezerocoords}
\end{equation}
we get the approximated line element
\begin{equation}
ds^2\simeq\frac{b^2}{\rho^2}(d\rho^2+d\theta^2)+r^2_+d\Sigma_N^2.
\label{doublezerometric}
\end{equation}

So, once the compactification in the Euclidean time is carried over, the
manifold shows the
topology~${\mathbb{H}}^2\!/\Gamma\times\Sigma^N$,~${\mathbb{H}}^2$ being
the
two-dimensional hyperbolic space, and~$\Gamma$
being the (discontinuous and fixed-point-free) group of isometry induced
by
the identification~$\theta\sim\theta+n\gamma$.

Notice that in this case it is not possible to determine the temperature
by
using the method of the conical singularity, since no conical singularity
is
present. It is anyway not correct to deduce from this fact
that such a manifold admits
any temperature~\cite{hawking2}, since it is
known~\cite{anderson,moretti0} 
that,~$T=0$ is the only physical temperature admissible in
the
case of a four-dimensional extreme Reissner-Nordstr\"om black hole (which
can
be recovered by setting~$d\Sigma_N^2=d\Omega_2$
in~(\ref{doublezerometric})).
Futher evidence in favour of this fact comes from the absence of the 
Hawking radiation in the extreme Reissner-Nordstr\"om black
hole~\cite{vanzo}.

Again we will see that this is the
only temperature which gives a well behaved~$\langle\phi^2(x)\rangle$ on
thehorizon, and a smooth manifold near it.\\

Without loss of generality, we set~$b^2$ as well as~$r_+$ equal to 1.

\section{First Case:
${\mathcal{M}}^D={\mathcal{C}}_\gamma\times\Sigma^N$.}

As we mentioned in the Introduction, our method requires the evaluation of
the
expectation value of the squared of the scalar field, using 
the~\mbox{$\zeta$-function} regularization tecnique.
We can so start by reviewing the evaluation of the heat kernel and the 
local~\mbox{$\zeta$-function} 
on~${\mathcal{M}}^D={\mathcal{C}}_\gamma\times\Sigma^N$
(for a complete discussion see~\cite{zerbini}).

We consider a massless and minimally coupled scalar field
on~${\mathcal{C}}_\gamma$, so that the associated operator is the pure
Laplacian~$L_\gamma=-\nabla_\gamma=\partial^2_\rho+\frac1{\rho}\partial_\rho+
\frac1{\rho^2}\partial^2_\theta$; the spectral properties of this operator
are
well known, and, in fact, a complete set of normalized eigenfunctions is
easily found to be
\begin{equation}
\psi_{n\lambda}=\frac1{\sqrt\gamma}e^{\frac{2\pi ni}{\gamma}\theta}
J_{\nu_n}(\lambda\rho), \qquad \nu_n=\frac{2\pi\vert n\vert}{\gamma},
\qquad n\in {\mathbb{Z}},
\end{equation}
together with its complex conjugate.\\
Here~$\lambda^2$~($\lambda\geq 0$) is the eigenvalue corresponding 
to~$\psi$ and~$\psi^*$, while~$J_\nu$ is the regular Bessel function.
So, using the standard separation of variables, it is easy to get the
spectrum
and eigenfunctions of the operator~$L_D=-\nabla_\gamma+L_N$
on~${\mathcal{M}}^D={\mathcal{C}}_\gamma\times\Sigma^N$,~$L_N$ being a
Laplace-like operator on~$\Sigma^N$ including, eventually, a mass and a
scalar
curvature coupling term. Moreover,
since we suppose~$\Sigma^N$ an arbitrary smooth manifold without boundary,
all
known results concerning the heat kernel and the~\mbox{$\zeta$-function}
for~$L_N$
on~$\Sigma^N$ (which we assume to be known) are applicable.\\
In particular the heat kernel has the usual asymptotic expansion (see also
Sec.~\ref{secfieldfluctuations})
\begin{equation}
K(t;\vec x\vert L_N)\simeq\sum_{r=0}^{\infty}A_r(\vec x\vert
L_N)t^{r-\frac N2},
\label{transversehk}
\end{equation}
and the meromorphic structure of the local~\mbox{$\zeta$-function} reads
\begin{equation}
\Gamma(s)\zeta(s;\vec x\vert L_N)=\sum_{r=0}^{\infty}\frac{A_r(\vec
x\vert L_N)}{s+r-\frac N2}+J(s;\vec x\vert L_N),
\label{transversezf}
\end{equation}
where~$J(s;\vec x\vert L_N)$ is the (generally unknown) analytic part. 
Here we have supposed the absence of zero modes, but one can easily take
them
into account with a simple modification of the formulas.

We can now derive the meromorphic structure of~$\zeta_\gamma(s;x\vert
L_D)$
on~${\mathcal{M}}^D={\mathcal{C}}_\gamma\times\Sigma^N$. To this aim, one
can use the factorization property of the heat kernel
\begin{equation}
K_\gamma(t;x\vert L_D)=K(t;{\theta,\rho}\vert L_\gamma)K(t;\vec x\vert
L_D),
\label{factorizedhk}
\end{equation}
in which the heat kernels of the Laplace-like operators
on~${\mathcal{M}}^D$,~${\mathcal{C}}_\gamma$ and ~$\Sigma^N$, respectively
appear.\\
By taking the Mellin transform of~(\ref{factorizedhk}), one usually gets
the
Dikii-Gelfand representation of the~\mbox{$\zeta$-function}, from which
the
meromorphic structure can be deduced.

Anyway in dealing with the conical manifold one has a convergence
obstruction,
in the meaning that there are no value of~$s$ for which the Mellin
transform
of~(\ref{factorizedhk}) is a finite quantity. The solution to this problem
has
been suggested by Cheeger~\cite{cheeger}, and simply consist in a
separation
between higher and lower eigenvalues. In practice we split the sum which
appears in the heat kernel (and in the related~\mbox{$\zeta$-function}) in
two sums,
the first over the lower eigenvalues, and the second over the higher ones;
then, after the analytic continuation is performed, one may define the 
full~\mbox{$\zeta$-function} by summing up the two
contributions obtained in this way (of course such a definition has all
the
requested properties and coincides with the usual one if the manifold is
smooth).\\
So we set
\begin{eqnarray}
\zeta_<(s;x\vert L_D) & = &
\int_0^\infty\!dt\,t^{s-1}K_<(t;{\theta,\rho}\vert
L_\gamma)K(t;\vec x\vert L_N), \\
\zeta_>(s;x\vert L_D) & = &
\int_0^{\infty}\!dt\,t^{s-1}K_>(t;{\theta,\rho}\vert
L_\gamma)K(t;\vec x\vert L_N),
\end{eqnarray}
where~$K_<(t;{\theta,\rho}\vert L_D)$ and~$K_>(t;{\theta,\rho}\vert L_D)$
are,
respectively, the ``lower'' and the ``higher'' heat kernels, which are
related
to the corresponding~\mbox{$\zeta$-function} by the relations
\begin{eqnarray}
K_<(t;{\theta,\rho}\vert L_\gamma) & = & \frac1{2\pi i}\int_{\frac12<
{\mathrm{Re}}\,(s)<1}\!ds\,t^{-s}\Gamma(s)\zeta_<(s;{\theta,\rho}\vert
L_\gamma),\\
K_>(t;{\theta,\rho}\vert L_\gamma) & = & \frac1{2\pi i}\int_{\frac12<
{\mathrm{Re}}\,(s)<1+\nu_1}\!ds\, t^{-s}\Gamma(s)\zeta_>(s;{\theta,\rho}
\vert L_\gamma), \\
\zeta_<(s;{\theta,\rho}\vert L_\gamma) & = &
\frac1{\Gamma(s)}\int_0^\infty
\!dt\,t^{s-1}K_<(t;{\theta,\rho}\vert L_\gamma),
\qquad \frac12<{\mathrm{Re}}\,(s)<1, \\
\zeta_>(s;{\theta,\rho}\vert L_\gamma) & = &
\frac1{\Gamma(s)}\int_0^\infty
\!dt\,t^{s-1}K_>(t;{\theta,\rho}\vert L_\gamma),
\qquad \frac12<{\mathrm{Re}}\,(s)<1+\nu_1,
\end{eqnarray}
and, by definition,
\begin{eqnarray}
K_\gamma(t;{\theta,\rho}\vert L_\gamma) & = &
K_<(t;{\theta,\rho}\vert L_\gamma) + K_>(t;{\theta,\rho}\vert L_\gamma),\\
\zeta_\gamma(s;{\theta,\rho}\vert L_\gamma) & = &
\zeta_<(s;{\theta,\rho}\vert L_\gamma)+\zeta_>(s;{\theta,\rho}\vert
L_\gamma).
\end{eqnarray}

Now, making use of  the Mellin-Parseval identity and paying attention to
the
range of convergence, one gets, for~$\textrm{Re}\,(s)>1+\frac N2$, the
following representation~\cite{zerbini}
\begin{equation}
\zeta_\gamma(s;x\vert L_D)\simeq
\frac{\zeta(s-1;\vec x\vert L_N)}{2\gamma
(s-1)}+\frac1{\gamma\Gamma(s)}\sum_{r=0}^PA_r(\vec x\vert L_N)I_\gamma
(s+r-\frac N2)\rho^{2s+2r-D}+{\mathcal{O}}(\rho^{2s+2P-D}),
\label{zetarep}
\end{equation}
where $P$ is an arbitrary large integer, while
\begin{equation}
I_\gamma(s)=\frac{\Gamma\left(s-\frac12\right)}{\sqrt
\pi}\left[G_\gamma(s)+
G_{2\pi}(s)\right].
\end{equation}
For~${\mathrm{Re}}\,(s)>1$, 
\begin{equation}
G_\gamma(s)=\sum_{n=1}^\infty\frac{\Gamma\left(\nu_n-s+1\right)}
{\Gamma\left(\nu_n+s\right)},
\qquad G_{2\pi}=-\frac{\Gamma(1-s)}{2\Gamma(s)}.
\end{equation}
It is possible to show that~$G_\gamma(s)$ admits an analytical
continuation,
and the properties of~$I_\gamma$ and~$G_\gamma$ on the whole complex plane
are
studied in detail in the Appendix of~\cite{zerbini}; an important property
is
that the analytical continued~$I_\gamma$ as well as~$G_\gamma$, has only a
simple pole at~$s=1$, with residue
\begin{equation}
{\mathrm{Res}}\,I_\gamma(s)_{\vert s=1}=\frac12\left(\frac{\gamma}{2\pi}-1
\right).
\label{residue}
\end{equation}

Having found the meromorphic structure of the~\mbox{$\zeta$-function} on
our 
manifold, we can
determine the vacuum expectation value of the fluctuation of a scalar
field.
With regard to this, it is convenient to distinguish 
between odd- and even-dimensional space-times.

We first consider the case in which~$N$ (or, equivalently,~$D$) is odd, so
that~$I_{\gamma}$ is finite at~$s=1$. To begin with, notice that 
the first term in~(\ref{zetarep})
depends only on the transverse coordinates and is finite on the horizon.
As a result, making use of the meromorphic structure
of~$\zeta(s;\vec x\vert L_N)$, we get
\begin{equation}
\langle\phi^2(x)\rangle\simeq
\frac1{2\gamma}\left[\sum_{r=0}^{\infty} 
\frac{A_r(\vec x\vert L_N)}{r-\frac N2}+J(0;\vec x\vert L_n)\right]+
\frac1{\gamma}\sum_{r=0}^PA_r(\vec x\vert L_N)
I_{\gamma}(1+r-\frac N2)\rho^{2r-N}+
{\mathcal{O}}(\rho^{2P-N}).
\end{equation}
It is now easy to see that the above expression
contains~$\left[\frac N2\right]$ (where~[ ] means ``integer part'')
terms which are divergent as~$\rho^{2r-N}$~($r<\left[\frac N2\right]$)
in the limit~$\rho\to0$, and so on the horizon
(see~(\ref{simplezerocoords})).
Thus, if we want a good behavior on
it, we must demand that all the~$I_{\gamma}$'s vanish
for~$r<\left[\frac N2\right]$, \emph{i.e.}~$\gamma=2\pi$. In particular
notice
that within this value of~$\gamma$ all of the~$I_{\gamma}$'s actually
vanish.

We now come to the case in which~$N$~($D$) is even. In this case,
the~\mbox{$\zeta$-function}~(\ref{zetarep}) has a  pole at~$s=1$, coming
from the
first and the~$I_\gamma$ term in~(\ref{zetarep}). 
From~(\ref{evenfieldfluctuations}) one gets
\begin{eqnarray}
\langle\phi^2(x)\rangle_{\mathrm{ren}} & \simeq & \frac1{2\gamma}
\left[ \sum_{r=0}^{\infty}\,'
\frac{A_r(\vec x\vert L_N)}{r-\frac N2}+J(0;\vec x\vert L_n) \right]
-\frac1{4 \pi}A_{\frac N2}(\vec x\vert L_N)\ln\mu^2 \nonumber \\
& & +\frac1{\gamma}\sum_{r=0}^P\,'
A_r(\vec x\vert L_N)I_\gamma(1+r-\frac N2)\rho^{2r-N}
+{\mathcal{O}}(\rho^{2P-N}),
\end{eqnarray}
where the $'$ in the sums means obmission of the~$r=\frac N2$ term.\\
Again, as long as~$r<\frac N2$, we get divergent terms on the horizon,
unless
we require the~$I_{\gamma}$'s to vanish, \emph{i.e.}~$\gamma=2\pi$ as in
the
odd-dimensional case.

As a result, for a manifold~${\mathcal{M}}^D$ whose near horizon geometry is
described by~${\mathcal{C}}_{\gamma}\times\Sigma^N$, the requirement of 
having a well behaved~$\langle\phi^2(x)\rangle$ on the horizon, 
selects~$\gamma=2\pi$,
and so the Unruh-Hawking temperature, according to the conical singularity 
method.\\
We also remind that the choice~$\gamma=2\pi$ makes the manifold smooth,
getting rid of the conical singularity otherwise present in~$\rho=0$.

The computation of the partition function for arbitrary $\gamma$
has be done in \cite{zerbini}, and it as be used in order to discuss
thermodinamical properties. Only the horizon divergences are present, and
these are still present in the on-shell ($\gamma=2\pi$) entropy.

\section{Second Case: ${\mathcal{M}}^D={\mathbb{H}}^2\!/\Gamma\times
\Sigma^N$.}

After having checked that our procedure works at least in the case of
non-extreme
black holes, we can tackle the case of the extreme ones, \emph{i.e.}
manifold
whose topology near the horizon is described by~${\mathbb{H}}^2\!/\Gamma
\times\Sigma^N$.

Again, one can start by making use of the factorization property of the
heat
kernel, writing that
\begin{equation}
K_\gamma(t;x\vert L_D)=K(t;{\theta,\rho}\vert L_\gamma)K(t;\vec x\vert
L_N),
\end{equation}
where the heat kernels of the Laplace-like operators
on~${\mathcal{M}}^D$,~${\mathbb{H}}^2\!/\Gamma$ and ~$\Sigma^N$,
respectively
appear.\\
As in the previous case, we suppose that~$L_\gamma$ is the operator
associated
to a massless and minimally coupled scalar field on~${\mathbb{H}}^2$,
while~$L_N$ is a Laplace-like operator on~$\Sigma^N$ including eventually,
mass and
scalar curvature coupling term (so that the
expansions~(\ref{transversehk})
and~(\ref{transversezf}) are still valid). In this way,~$L_\gamma=
-\Delta_\gamma=-\rho^2(\partial^2_\theta+\partial^2_\rho)$, and a complete
set
of normalized eigenfunctions is easily found to be
\begin{equation}
\psi_{\lambda k}=\sqrt{\frac y{2\pi}}e^{ik\theta}K_{i\lambda}(\vert k\vert
y),
\end{equation}
together with its complex conjugate.\\
Here~$\lambda^2$~($\lambda\geq 0$) is the eigenvalue corresponding
to~$\psi$
and~$\psi^*$, while~$K_\nu$ is the MacDonald function.
Thus the spectral representation of the (off-diagonal) heat kernel
associated
to~$L_\gamma$ on~${\mathbb{H}}^2$ reads (see, for example~\cite{bytsenko})
\begin{equation}
K_{{\mathbb{H}}^2}(t;{\theta,\rho;\theta',\rho'}\vert
L_\gamma)=\frac1{2\pi}
\int_0^{\infty}\!d\lambda\, \lambda\tanh(\pi\lambda)
e^{-\left(\lambda^2+\frac14\right)t}P_{i\lambda-\frac12}(\cosh \sigma),
\end{equation}
where~$P$ is the associated Legendre function, while~$\sigma$ is
the~${\mathbb{H}}^2$ geodesic distance between~$({\theta,\rho})$
and~$({\theta',\rho'})$.

As previously remarked, for studying thermal effects, one has to deal with
the
quotient space~${\mathbb{H}}^2\!/\Gamma$, with~$\Gamma$ the (discontinuous
and
fixed-point-free) group of isometry induced by the time
compactification~$0\leq\theta\leq\gamma$. In our case we have traslations,
corresponding to parabolic elements. By applying the method of images, the
diagonal heat kernel turns out to be
\begin{eqnarray}
K(t;{\theta,\rho}\vert L_\gamma) & = & \frac1{2\pi}\int_0^{\infty}\!
d\lambda\,\lambda\tanh(\pi\lambda)e^{-\left(\lambda^2+\frac14\right)t} 
\nonumber \\
& & +\frac1{\pi}\sum_{n=1}^{\infty}\int_0^{\infty}\!d\lambda\,\lambda
\tanh(\pi\lambda)e^{-\left(\lambda^2+\frac14\right)t}P_{i\lambda-\frac12}
(\cosh\sigma_n),
\end{eqnarray}
where now
\begin{equation}
\cosh\sigma_n=1+\frac{n^2\gamma^2}{2\rho^2}.
\end{equation}
Let us show that the partition function does not exist, and requires,
besides
the horizon divergence regularization, a further regularization. The
partition
function is proportional to the first derivative of
the~\mbox{$\zeta$-function}, which may be defined by the Mellin transform
of
the heat kernel trace. The latter may be obtained integrating over the
manifold coordinates. As a result
\begin{eqnarray}
\zeta\left(s\vert L_\gamma\right) & = & \frac{\gamma}{4\pi\varepsilon}
\frac1{s-1}\zeta(1-s\vert L_N+\frac14)+ \frac{\gamma}{2\pi\varepsilon}
\int_0^\infty d\lambda\frac{\lambda}{1+e^{2\pi\lambda}}
\zeta(s\vert L_N+\lambda^2+\frac14) \nonumber \\
& & +\frac1{s\sqrt\pi\gamma^{\delta}}\zeta_{\mathrm{R}}(1+\delta)
\zeta(s-\frac12\vert L_N+\frac14)+{\mathcal{O}}(\delta),
\end{eqnarray}
where $\zeta_{\mathrm{R}}$ is the Riemann~\mbox{$\zeta$-function}, and 
we have introduced the horizon cutoff $\varepsilon$, and the cusp
regularization $\delta>0$. It should be noticed the divergence for
$\delta=0$,
which is usually present when one is dealing with parabolic elements
\cite{binosi}.

As far as the field fluctuations are concerned, we only need the
expression of
the local $\zeta$-function near the horizon. Thus with regard to 
the sum over~$n$, we may apply the simplest version of
the Euler-MacLaurin resummation formula, namely
\begin{equation}
\sum_{n=1}^{\infty}f(n)=\int_1^{\infty}\!dx\, f(x)-\frac12f(1)+
\int_1^{\infty}\!dx\,\left(x-[x]-\frac12\right)f'(x).
\end{equation}
As a result, for large~$\rho$
\begin{equation}
\sum_{n=1}^{\infty}P_{i\lambda-\frac12}(\cosh\sigma_n)=
\frac{\rho}{\sqrt 2 \gamma}\frac1{\lambda\tanh(\pi\lambda)}+
C(\lambda)+{\mathcal{O}}(\frac{\gamma}{\rho}),
\label{emlresummation}
\end{equation}
where~$C(\lambda)$ does not depend on~$\rho$.\\
Thus, the diagonal part of the heat-kernel may be rewritten as
\begin{equation}
K(t;{\theta,\rho}\vert L_\gamma)=\frac1{2\pi}\int_0^{\infty}\!
d\lambda\,\lambda\tanh(\pi\lambda)e^{-\left(\lambda^2+\frac14\right)t}
+\frac{\rho}{2\gamma\sqrt{2\pi t}} e^{-\frac{t}{4}}+{\mathcal{O}}(1)+
{\mathcal{O}}(\frac{\gamma}{\rho}).
\end{equation}
As a result, the related local~\mbox{$\zeta$-function} reads
\begin{equation}
\zeta_\gamma(s;x\vert L_D)=\zeta(s;\vec x\vert L_D)+
\frac{\rho}{2\gamma\sqrt{2\pi}}
\frac{\Gamma\left(s-\frac{1}{2}\right)}{\Gamma(s)}
\zeta(s-\frac{1}{2};\vec x\vert L_N+\frac{1}{4})
+{\mathcal{O}}(1)+{\mathcal{O}}(\frac{\gamma}{\rho}).
\label{mmm}
\end{equation}
The first term on the r.h.s. of the above relation, is the 
local~\mbox{$\zeta$-function} associated with
the smooth manifold~${\mathcal{M}^D}={\mathbb{H}}^2 \times\Sigma^N$
and it depends only on the transverse coordinates~$\vec x$. The other
terms
are the asymptotic contribution for large~$\rho$. 
As a result, we have obtained the meromorphic structure
of~$\zeta(s;x\vert L_D)$ via the meromorphic structure
of~$\zeta_\gamma(s;\vec x\vert L_D)$
and~$\zeta(s-\frac{1}{2};\vec x\vert L_N+\frac{1}{4})$. It should be
noticed
that the second term,  which is the relic of the sum over images, contains  
the shift~$s-1/2$. This means that, with regard to the evaluation of 
the field fluctuation, one has a simple pole at~$s=1$ for any~$D$, not
only
for~$D$ even. This violation of Seeley's Theorem is related to the
presence of 
parabolic elements, which make the manifold not smooth.  

On the other side, no matter the dimension of the space-time,
the field fluctuations contain terms proportional to~$\frac{\rho}{\gamma}$
which are divergent on the horizon~($\rho\to\infty$,
see~(\ref{doublezerocoords})), unless we demand~$\gamma=\infty$, and
so~$T=0$,
according to the result obtained in~\cite{anderson,moretti0} 
in the four-dimensional case. We finally notice that if $\gamma=\infty$,
the
partition function contains only the usual volume divergence associated to
the
non-compact nature of the Euclidean section.

\section{Conclusions}

In this paper, making use of an Euclidean approach, we have evaluated the
field fluctuations and the partition function of a scalar field in a
$D$-dimensional static black hole. A near horizon approximation has lead
to
quite explicit expressions for the local~\mbox{$\zeta$-function} related
to
such a field propagating in the Euclidean section of the black hole
space-time. The period of the compactified imaginary time has been
interpreted as the inverse of the temperature; further, making use
of~\mbox{$\zeta$-function} regularization, the field fluctuations have
been
evaluated. It is then stressed the fact that this quantity is finite on
the
horizon as soon as one selects a distinguished period of the imaginary
time
(temperature), which coincides with the Hawking-Unruh temperature in the
non-extreme case, and with the zero temperature in the extreme one.
With regard to the partition function, no problem exists in the conical
singularity case, while in the presence of the cusp singularity,
new divergences
show up in the global~\mbox{$\zeta$-function}, and, strictly speaking, the
partition function does not exist. This drawback disappears if the cusp
singularity is absent, namely if the temperature is again zero.

With regard to the local quantities, we also notice that the regular
behavior
of the fluctuations on the horizon, leads also to the regular behavior of
the
expectation value of the stress tensor. This can be verified by means of a
direct calculation, starting from the local
off-diagonal~\mbox{$\zeta$-function} which can be obtained with our
approach.

As far as the extreme black holes are concerned, all the properties we
have
been deriving, and the lack of the Hawking radiation \cite{vanzo},
strongly
suggest that the only admissible temperature is the zero one, in agreement
with the 4-dimensional case studied in \cite{anderson}.
                                                      
The only class of space-times for which our analysis seems to have no
direct 
application is the one in which the double zero  occurs at~$r_+=0$. 
As an example, we may recall the so called massless ground state of the 
asymptotically AdS toroidal black
holes~\cite{vanzo1,mann,brill,birmingham}. In fact these black holes have 
\begin{equation}
f(r)=\frac{1}{h(r)}=\left(\frac{l^2}{r^2}-\frac{C_D M}{r^{D-1}} \right)\,,
\end{equation}     
where~$C_D$ is a constant,~$M$ is the mass of the black hole and 
the parameter~$l$ is related to the 
cosmological constant, namely~$\Lambda=-l^{-2}$. For~$D=3$, one recovers 
the celebrated BTZ black hole~\cite{banados}. The ground state of this
class
of black holes is the zero mass solution, and the Euclidean metric becomes
\begin{equation}
ds^2=\frac{r^2}{l^2}d\tau^2+\frac{l^2}{r^2}dr^2+r^2dT_N^2,
\end{equation}
where~$dT_N^2$ represents the metric of a $N$-dimensional torus. 
In the above metric,~$r=0$ is a naked coordinate singularity.
If one compactifies  the Euclidean time and make the coordinate 
transformation~$r=l^2/\rho$, one gets
\begin{equation}
ds^2=\frac{l^2}{\rho^2}\left[d\rho^2+d\tau^2+l^2dT_N^2\right].
\end{equation}
This metric describes locally the~$D$-dimensional hyperbolic
space~${\mathbb{H}}^D$.\\
For the zero temperature case and in~$D=3$ and $D=4$, the field
fluctuations
as well as the expectation value of the stress tensor has been computed 
in~\cite{lyfschitz,binosi} and~\cite{caldarelli} respectively,
and divergences have been found as~$\rho$ goes to infinity.
In this case, our analysis does not select any distinguished temperature.
However, it should be noticed
that in this case, it is not reasonable to neglect the back-reaction
effects. In fact, in~\cite{lyfschitz,binosi,caldarelli} it has been
shown that there is a quantum implementation of the Cosmic Censorship
Principle due to the back-reaction on the metric.  

\section*{Aknowledgments.}

Useful discussions with V.~Moretti and L.~Vanzo are gratefully
aknowledged.

\end{document}